\documentclass[twocolumn]{aastex631}

\shorttitle{High Frequency Waves in Chromospheric Spicules}
\shortauthors{Bate et al.}
\graphicspath{{./}{figures/}}

\usepackage{gensymb}

\begin{document}

\title{High Frequency Waves in Chromospheric Spicules}

\correspondingauthor{W.~Bate}
\email{wbate02@qub.ac.uk}

\author[0000-0001-9629-5250]{W.~Bate}
\affiliation{Astrophysics Research Centre, School of Mathematics and Physics, Queen's University Belfast, Belfast, BT7 1NN, UK}

\author[0000-0002-9155-8039]{D.~B.~Jess}
\affiliation{Astrophysics Research Centre, School of Mathematics and Physics, Queen's University Belfast, Belfast, BT7 1NN, UK}
\affiliation{Department of Physics and Astronomy, California State University Northridge, 18111 Nordhoff Street, Northridge, CA 91330, USA}

\author[0000-0001-6423-8286]{V.~M.~Nakariakov}
\affiliation{Centre for Fusion, Space and Astrophysics, Physics Department, University of Warwick, Coventry, CV4 7AL, UK}
\affiliation{St. Petersburg Branch, Special Astrophysical Observatory, Russian Academy of Sciences, 196140, St. Petersburg, Russia}

\author[0000-0001-5170-9747]{S.~D.~T.~Grant}
\affiliation{Astrophysics Research Centre, School of Mathematics and Physics, Queen's University Belfast, Belfast, BT7 1NN, UK}

\author[0000-0002-7711-5397]{S. Jafarzadeh}
\affiliation{Leibniz Institute for Solar Physics (KIS), Sch{\"{o}}neckstr. 6, 79104 Freiburg, Germany}
\affiliation{Rosseland Centre for Solar Physics, University of Oslo, P.O. Box 1029 Blindern, NO-0315 Oslo, Norway}

\author[0000-0002-5365-7546]{M. Stangalini}
\affiliation{ASI, Italian Space Agency, Via del Politecnico snc, 00133, Rome, Italy}

\author[0000-0001-8556-470X]{P.~H. Keys}
\affiliation{Astrophysics Research Centre, School of Mathematics and Physics, Queen's University Belfast, Belfast, BT7 1NN, UK}

\author[0000-0003-1746-3020]{D.~J. Christian}
\affiliation{Department of Physics and Astronomy, California State University Northridge, 18111 Nordhoff Street, Northridge, CA 91330, USA}

\author[0000-0001-5435-1170]{F.~P. Keenan}
\affiliation{Astrophysics Research Centre, School of Mathematics and Physics, Queen's University Belfast, Belfast, BT7 1NN, UK}



\begin{abstract}

Using high cadence observations from the Hydrogen-alpha Rapid Dynamics camera imaging system on the Dunn Solar Telescope, we present an investigation of the statistical properties of transverse oscillations in spicules captured above the solar limb. At five equally separated atmospheric heights, spanning approximately $4900-7500$~km, we have detected a total of $15{\,}959$ individual wave events, with a mean displacement amplitude of $151\pm 124$~km, a mean period of $54\pm 45$~s, and a mean projected velocity amplitude of $21\pm 13$~km{\,}s$^{-1}$. We find that both the displacement and velocity amplitudes increase with height above the solar limb, ranging from $132\pm 111$~km and $17.7\pm 10.6$~km{\,}s$^{-1}$ at $\approx4900$~km, and $168\pm 125$~km and $26.3\pm 14.1$~km{\,}s$^{-1}$ at $\approx7500$~km, respectively. Following the examination of neighboring oscillations in time and space, we find 45\% of the waves to be upwardly propagating, 49\% to be downwardly propagating, and 6\% to be standing, with mean absolute phase velocities for the propagating waves on the order of $75-150$~km{\,}s$^{-1}$. While the energy flux of the waves propagating downwards does not appear to depend on height, we find the energy flux of the upwardly propagating waves decreases with atmospheric height at a rate of $-13{\,}200\pm6500$~W{\,}m$^{-2}$/Mm. As a result, this decrease in energy flux as the waves propagate upwards may provide significant thermal input into the local plasma.
\end{abstract}



\section{Introduction} 
\label{sec:intro}
Spicules are dynamic plasma jets that are prevalent within the solar chromosphere, and which generally have diameters on the order of hundreds of km. They are relatively short-lived features, typically having a lifetime of less than $10$ minutes \citep{Pereira2012}. When viewed in the visible and UV bands at the solar limb, spicules appear ubiquitously as a dense forest of narrow, straw-like features \citep{1998ESASP.421...35S, Sterling2000}. 

\begin{figure*}[ht!]
\includegraphics[trim=0mm 0mm 0mm 0mm, clip, width=\textwidth, angle=0]{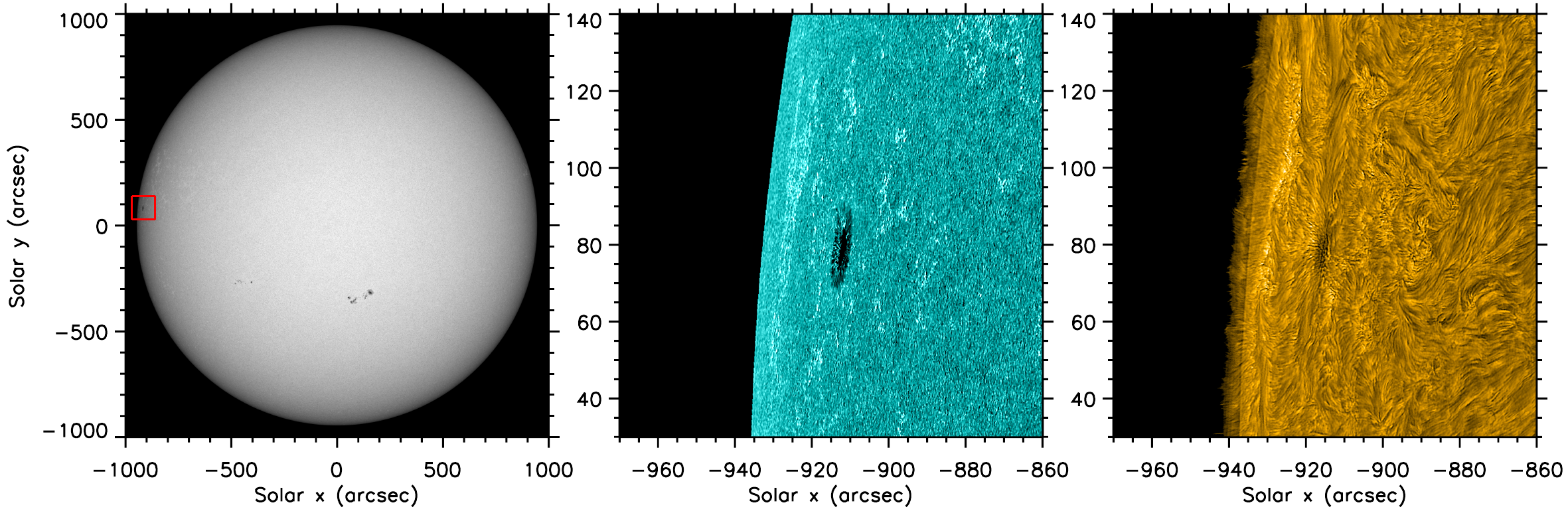}
 \caption{Contextual SDO/HMI continuum (left), ROSA G-band (middle), and H$\alpha$ line core (right) images acquired at 13:59:09~UT. The area imaged by ROSA and HARDcam is marked by the red square in the full disk SDO/HMI continuum image. Numerous spicules are clearly visible above the solar limb as narrow, straw-like structures in the corresponding H$\alpha$ image.
\label{fig:FOV}}
\end{figure*}

\citet{Secchi1877} was the first to observe solar spicules, and they have remained a focal point of solar physics research over the last 140 years. Transverse oscillations in spicules were first identified in the $1960$s \citep{Pasachoff1968}, utilizing ground-based observations obtained at the Sacramento Peak Observatory \citep[for a comprehensive review see][]{2009SSRv..149..355Z}. The magnetic cylinder model is generally accepted as being the most applicable to spicules, allowing their oscillatory behavior to be interpreted and modeled as magnetohydrodynamic \citep[MHD;][]{Alfven1942} modes \citep{Edwin1983}. \citet{Sterling2000} highlighted that high-resolution observations, due to the small width of the structures, are vital for a complete description of the spicule wave phenomena. \citet{Wedemeyer2007} also note that the ability to detect oscillatory power at higher frequencies is influenced by the spatial resolution of the observations \citep[see also the discussions provided by][]{2020NatAs...4..220J, 2021NatAs...5....5J}.

One of the major focuses of current solar physics research is the so-called `coronal heating paradox'. One of the proposed theoretical mechanisms to explain the source of this heating is linked to the propagation and dissipation of wave phenomena, commonly referred to as the `AC' heating mechanism \citep{Alfven1947}. Spicules are of particular interest when attempting to explain the heating of the solar atmosphere due to their potential to facilitate the transfer of mass and energy between the photosphere and corona. They are often categorized by their properties into two types, type~{\sc{i}} and type~{\sc{ii}} \citep{DePontieu2007a, Pereira2012}, although such distinct classifications are still under debate \citep[e.g.,][]{Zhang2012, 2013ApJ...764...69P}.

Observations of transverse oscillations of spicules, fibrils, and mottles in the upper chromosphere typically find mean periods on the order of $80-300$~s, often with $\sim50-1000$ examples found during the course of an individual data sequence \citep{,1967SvA....10..744N,1971SoPh...18..403N,DePontieu2007, Kuridze2012, Morton2012, Morton2013, Morton2014}. A major exception are the short period ($45\pm30$~s) transverse oscillations found in $89$ type~{\sc{ii}} spicules within a coronal hole identified by \citet{Okamoto2011}. However, \cite{Okamoto2011} suggest that this short average period is likely due to them utilizing a methodology which did not allow for the measurement of the properties of the longer period waves ($>100$~s). Another exception is the more recent work of \citet{2021ApJ...921...30S}, which found 30 examples of transverse spicule oscillations with periods ranging from $16-100$~s. These authors also note a selection effect due to only choosing an event if its oscillation period is less than its lifetime, resulting in longer period waves not being considered.

Finding the energy flux of spicule oscillations is an important step in investigating their contribution to the heating of the chromosphere and corona. It is estimated that an energy flux of $10^{3}-10^{4}$~W{\,}m$^{-2}$ is required to heat the chromosphere. The energy required to heat the corona is around an order of magnitude less than that required to heat the chromosphere \citep{Withbroe1977}. This suggests that accounting for chromospheric heating is a challenge of equal or greater magnitude than for coronal heating when investigating solar atmospheric heating mechanisms.

Energy flux estimations are based on the interpretation of these transverse oscillations as MHD wave modes. \citet{DePontieu2007} interpreted the transverse oscillations of spicules as bulk Alfv{\'{e}}n waves and assumed a filling factor of unity. Using this interpretation, an energy flux estimate of $4000-7000$~W{\,}m$^{-2}$ was suggested. However, this bulk Alfv{\'{e}}n interpretation has attracted criticism, with \citet{2007Sci...318.1572E} and \citet{2008ApJ...676L..73V} pointing out that Alfv{\'{e}}n waves do not result in the bulk transverse motions observed, and instead proposing that these transverse oscillations are best interpreted as kink modes. The filling factor, a measure of what fraction of the total volume is occupied by oscillating spicules, is another extremely important consideration for energy flux calculations \citep{VanDoorsselaere2014}. An equivalent interpretation, assuming that spicules have an approximately constant width across varying heights, would be the ratio of the area of the solar surface covered by spicules to the total solar surface area.

\citet{2003PNAOJ...7....1M} found a spicule filling factor of $5\%$ at a height of $4000$~km using Ca~{\sc{ii}} H \& K line observations taken during a solar eclipse. This suggests that a filling factor of $0.05$ is more appropriate than unity. Using the revised interpretations of the most realistic MHD mode and associated filling factor, \citet{VanDoorsselaere2014} found that the energy flux estimates by \citet{DePontieu2007} were reduced from $4000-7000$~W{\,}m$^{-2}$ to $200-700$~W{\,}m$^{-2}$, a difference exceeding one order of magnitude. Furthermore, \citet{Morton2012} used high-resolution H$\alpha$ observations taken by the Dunn Solar Telescope to find a similar upper limit for the filling factor ($4-5$\%) for open chromospheric structures that connected to higher layers of the solar atmosphere. By interpreting the transverse oscillations of fibrils as kink modes, the authors estimated the energy flux as $170\pm110$~W{\,}m$^{-2}$, similar to that derived by \citet{VanDoorsselaere2014}. However, in addition to the $4-5$\% filling factors commonly used in modern literature, lower estimates have also been put forward, with \citet{1972ARA&A..10...73B} suggesting a filling factor of $0.6\%$. As a result, it is generally believed that the spicule filling factor spans an approximate order-of-magnitude (ranging between $\approx 0.5-5$\%), with differing values being applicable depending on factors such as the atmospheric height sampled and the degree of solar activity (i.e., it is not a quantity that can be applied universally across all observations). The influence of the chosen filling factor on energy flux calculations is discussed further in Section~\ref{sec:ana}.

An important caveat when interpreting these energy flux estimates is that they are only based on resolved transverse oscillations. Waves with amplitudes too small to be spatially resolved or periods too short to be temporally resolved are not included in these estimations, leaving the possibility that a significant amount of wave energy may be unaccounted for \citep{2016GMS...216..431V}. Another aspect contributing to the underestimation of the total energy flux may be the presence of kink motions along the observer's line-of-sight, which will not manifest as visible transverse oscillations. Examples of this have been documented by \citet{2018ApJ...853...61S} and \citet{2021ApJ...921...30S}, who measured helical motions of spicules through Doppler measurements \citep[see also the modeling work by][]{2008ApJ...683L..91Z}. If these line-of-sight motions are not taken into account when calculating the energy flux, it may result in an underestimation of the true value.

The aim of the current study is to identify the properties of spicule oscillations across a statistically significant sample that is extracted from different chromospheric heights. With oscillation characteristics measured across a range of atmospheric layers, we calculate the energy flux carried by these waves as a function of geometric height. To achieve this objective, we utilize ground-based instrumentation with high spatial and temporal resolutions, providing unprecedented data products that are ideally suited for this study.

\section{Observations} \label{sec:obs}
Our analysis employs data collected on 2015 July 27 from 13:52 -- 15:29~UT using the Dunn Solar Telescope \citep[DST;][]{Dunn1969} at the National Solar Observatory in New Mexico, USA. The Rapid Oscillations in the Solar Atmosphere \citep[ROSA;][]{2010SoPh..261..363J} and Hydrogen-alpha Rapid Dynamics camera \citep[HARDcam;][]{2012ApJ...757..160J} imaging systems were used to observe a large sunspot, which was part of NOAA~AR12391, close to the solar limb at N07.8E73.6 in the conventional heliographic coordinate system. Seeing conditions remained excellent throughout the first hour of the observing period, gradually worsening towards the latter stages of the observing window. 

HARDcam observations employed a narrow ($0.25$\AA~FWHM) bandpass filter centered on the H$\alpha$ line core ($6562.8$\AA), while the ROSA camera system observed the same region through G-band (10\AA~FWHM centered at 4305\AA) and broadband 4170{\AA} continuum filters. The HARDcam data have a pixel size of $0.092''$ ($66.5$~km), providing a $180''\times180''$ field-of-view, while the ROSA system was slightly undersampled ($0.180''$ per pixel) to provide an identical field-of-view size to that of the HARDcam observations. To correct for wavefront deformations in real time, higher-order adaptive optics (AO) were used during the observations \citep{Rimmele2004, Rimmele2011}. Original data from both ROSA and HARDcam were taken at a frame rate of $30.3$~s$^{-1}$, with the images synchronized by way of a master trigger with microsecond precision \citep{2010SoPh..261..363J}. The resulting HARDcam H$\alpha$ images were then improved using speckle reconstruction algorithms \citep{Woger2008}, utilizing a $30\rightarrow1$ restoration, resulting in a final reconstructed cadence of $0.990$~s. 

\begin{figure}[t]
\centering
  \includegraphics[trim=0mm 0mm 0mm 0mm, clip, width=\columnwidth, angle=0]{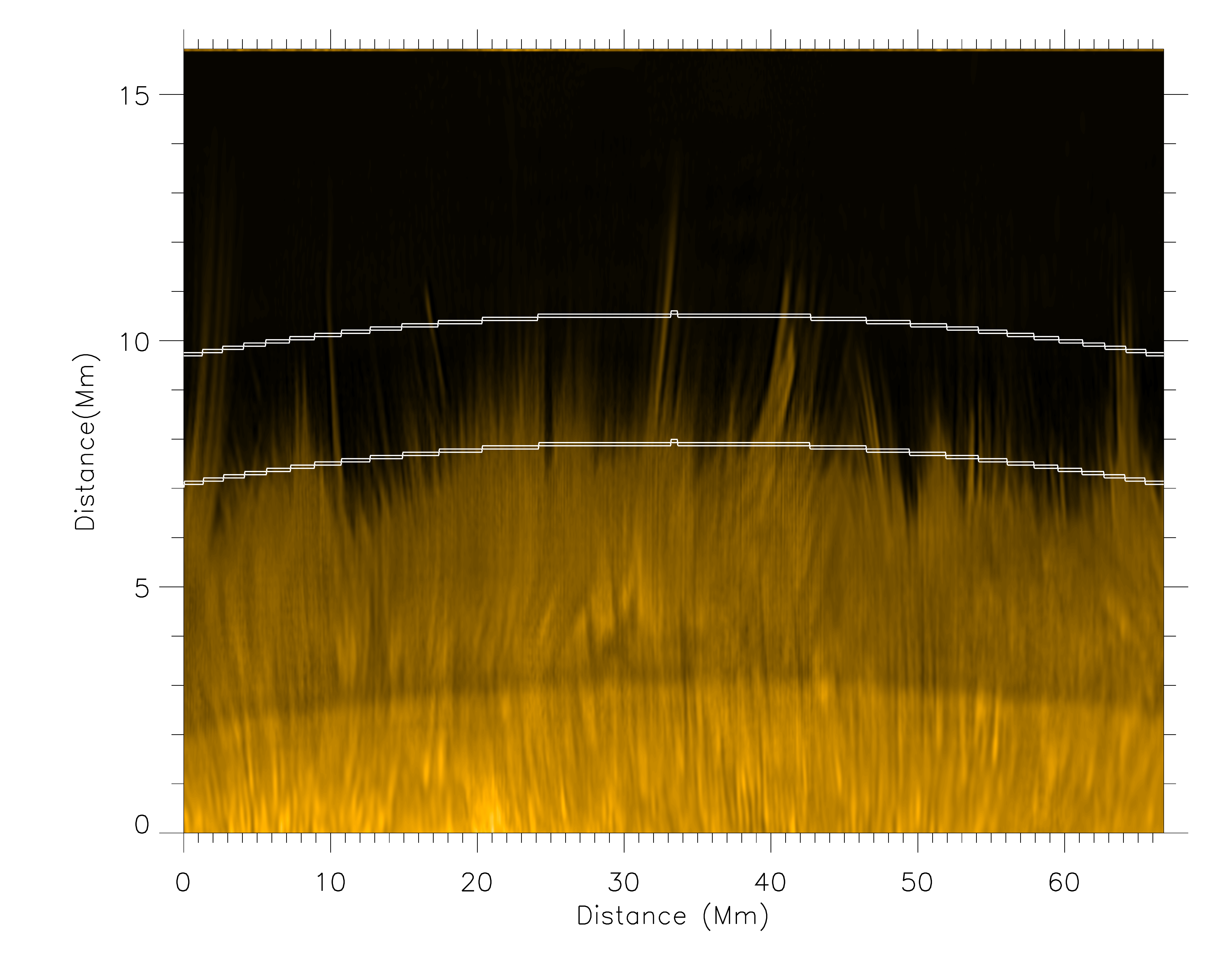}
\caption{An H$\alpha$ core sub-field ($67 \times 16$ Mm$^{2}$) image acquired using HARDcam at 14:49:45~UT. Numerous spicules are clearly visible above the solar limb as narrow, straw-like structures. The two most extreme slits used to take the time distance diagrams are shown by the white lines, at heights of $4890$ and $7500$~km. The axes are shown using different scales to aid with visual clarity.} 
\label{fig:slit_fov}
\end{figure}

ROSA continuum observations were coaligned using cross-correlation techniques \citep[see, e.g.,][]{2010ApJ...719L.134J} with contemporaneous continuum images from the Helioseismic and Magnetic Imager \citep[HMI;][]{Schou2012} on board the Solar Dynamics Observatory \citep[SDO;][]{Pesnell2012}, providing sub-arcsecond pointing accuracy for the field-of-view covered by the DST. Following this, the HARDcam field was aligned with the master ROSA images using sequences of targets acquired during the calibration procedures at the DST, resulting in H$\alpha$ observations that have precise pointing metadata that is consistent with modern space-based observatories. Contextual images from SDO/HMI, ROSA, and HARDcam, following the processing steps outlined above, are shown in Figure~{\ref{fig:FOV}}.

\section{Analysis and Discussion} 
\label{sec:ana}

During the course of the observations, the DST's AO system was locked onto the high-contrast sunspot structure that was very close to the limb. As a result, limb spicules close to the central portion of the field-of-view were accurately corrected from atmospheric seeing effects by the AO. Hence, the current HARDcam H$\alpha$ dataset offers an unprecedented opportunity to examine limb spicules at extremely high time cadence (0.990~s) and spatial resolution (133~km two-pixel resolution). 

\begin{figure}[t]
\centering
  \includegraphics[trim=0mm 0mm 0mm 0mm, clip, width=\columnwidth, angle=0]{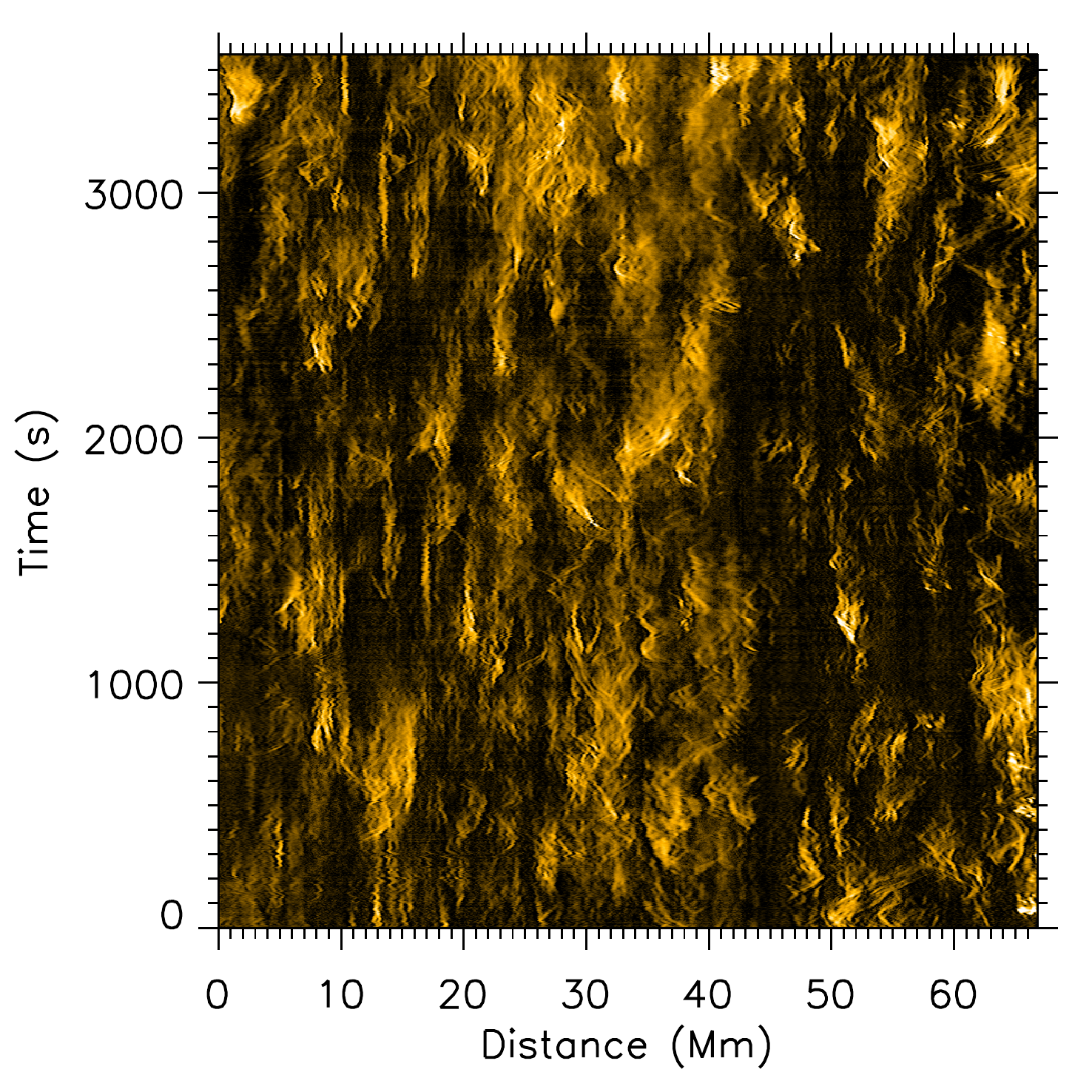}
\caption{A time-distance diagram captured using a curved slit at a height of $6850$~km above the solar limb. Each bright streak is a feature passing through the slit, with the clear oscillatory features representative of transverse motions displaying a range of amplitudes and periods.}
\label{fig:td}
\end{figure}

A sub-field, spanning approximately 70~Mm along the central portion of the field-of-view, where the AO corrections were operating optimally, was isolated for further study. As the DST was tracking the sunspot contained within the field-of-view, over time the pixel coordinates corresponding to the limb position change as a result of the sunspot rotating on to the disk. The image sequence was hence stabilized with respect to the limb, which was achieved by first choosing a reference frame towards the beginning of the dataset. Next, a small area of the limb image with high contrast was selected, with subsequent images compared and shifted using two-dimensional cross correlation techniques. Pixel shift values that produced the highest cross correlation coefficients were selected and applied to each image in the time series iteratively. The resulting shifted images lead to the limb remaining stationary at the same pixel location throughout the dataset, providing a robust baseline from which to examine spicule oscillations above the fixed limb.

\begin{figure}[t]
\centering
  \includegraphics[trim=8mm 12mm 5mm 8mm, clip, width=\columnwidth, angle=0]{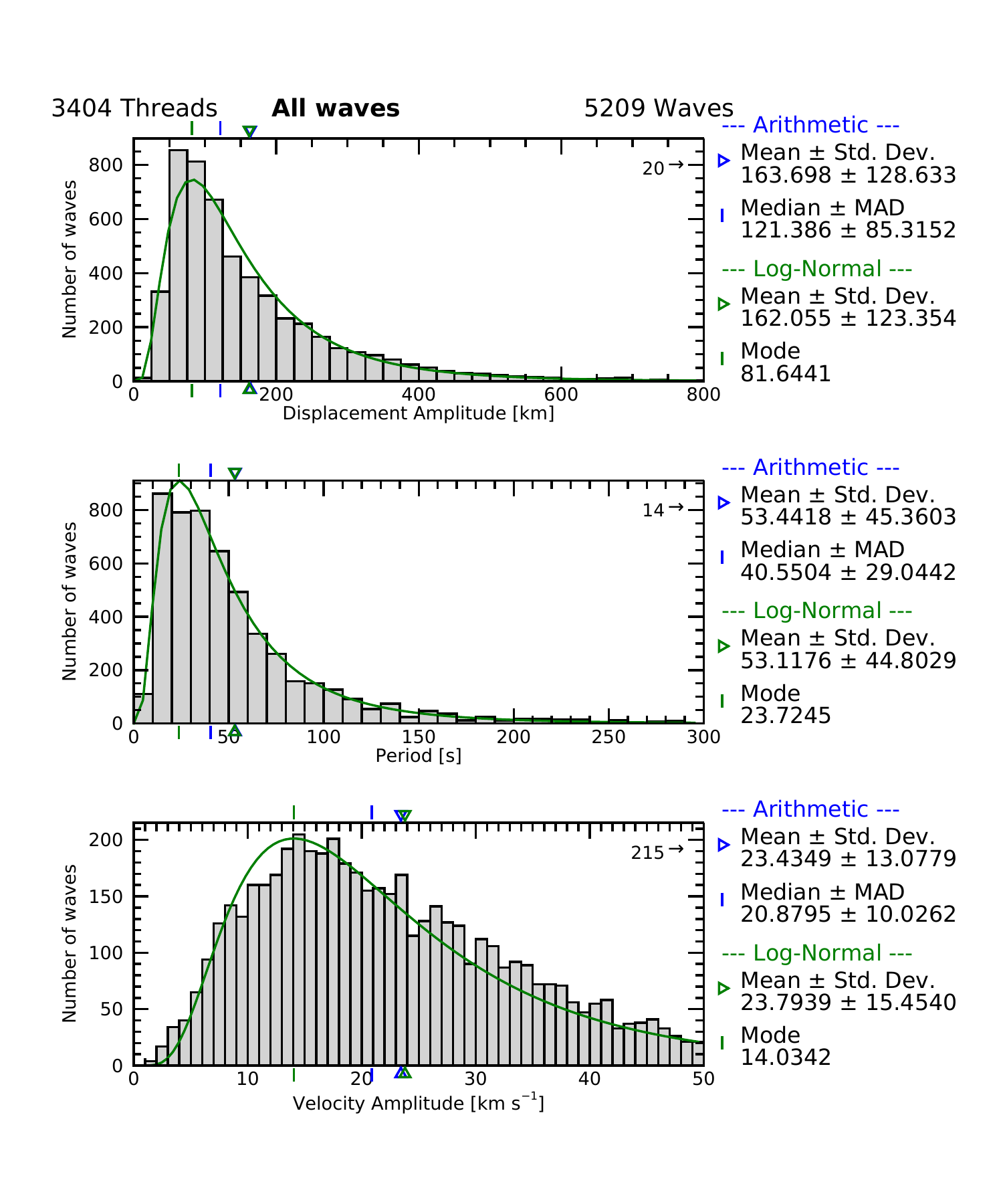}
\caption{Histograms of the wave properties identified at a height of $6850$~km above the solar limb. The upper, middle, and lower panels display information related to the displacement amplitudes, periods, and velocity amplitudes, respectively. Measurements of the corresponding averages and deviations are displayed on the right of each histogram. MAD denotes the median absolute deviation.}
\label{fig:histo}
\end{figure}

Multiscale Gaussian Normalization \citep[MGN;][]{Morgan2014} was applied to each image in the dataset in order to more easily identify each spicule and its associated motion. It must be noted that MGN does not preserve photometric accuracy. However, this is not an issue when mapping the transverse oscillations of features since we are not concerned with comparisons of relative intensities. For the application of MGN, we employed the convolution of HARDcam images with Gaussian kernels with one-sigma widths of $w = 1.25, 2.5, 5, 10, 20, 40$~pixels, followed by the production of gamma-transformed images with a $\gamma$ value of $3.2$ \citep{Poynton2003}.

Five slits were placed at equally spaced, constant radial heights above the limb, spanning approximately $4900$~km to $7500$~km in steps of $\approx650$~km. These slits were curved in nature in order to maintain a constant radial height above the limb, and the highest and lowest slits are shown by the white lines in Figure~\ref{fig:slit_fov}. When taking this approach, it is important to note that superposition along the line of sight of spicules anchored behind the limb, in front of the limb, and on the limb is unavoidable. As a result of the slit heights being based on a geometric distance above the limb, this will result in the foreground/background spicules being sampled further along their lengths than those precisely located on the limb. We have carefully selected the minimum and maximum heights of the slits to be in the range of $4890 - 7500$~km (see Figure~{\ref{fig:slit_fov}}), which is towards the upper end of the `dense forest' of spicules, hence minimizing the degree of feature superposition. Due to the (minimized) spicule superposition affecting each of the slits in a similar way, and considering the large numbers of spicules observed at each height, comparisons between wave properties taken with different slits will still be valid. However, it is still important to consider this effect when examining wave properties taken from a single slit in isolation, since the chosen slit height will be a minimum value of the distance sampled along the spicule due to these geometric considerations. Time-distance diagrams were then produced from each of the slits, with an example shown in Figure~\ref{fig:td}.

The Automatic Northumbria University Wave Tracking \citep[Auto-NUWT;][]{Morton2013, Weberg2018} code was utilized in order to identify the location of the spicules as a function of time, track their transverse motion, and extract the properties of their oscillations. Features are identified by fitting a sum of Gaussians to each time slice in the time-distance diagrams, enabling the determination of sub-pixel values for the location of the center of the feature. The transverse oscillatory behavior of these features is probed through the application of Fourier analysis to the position of the center of the feature as a function of time.

\begin{figure}[t]
\centering
  \includegraphics[trim=0mm 0mm 0mm 0mm, clip, width=\columnwidth, angle=0]{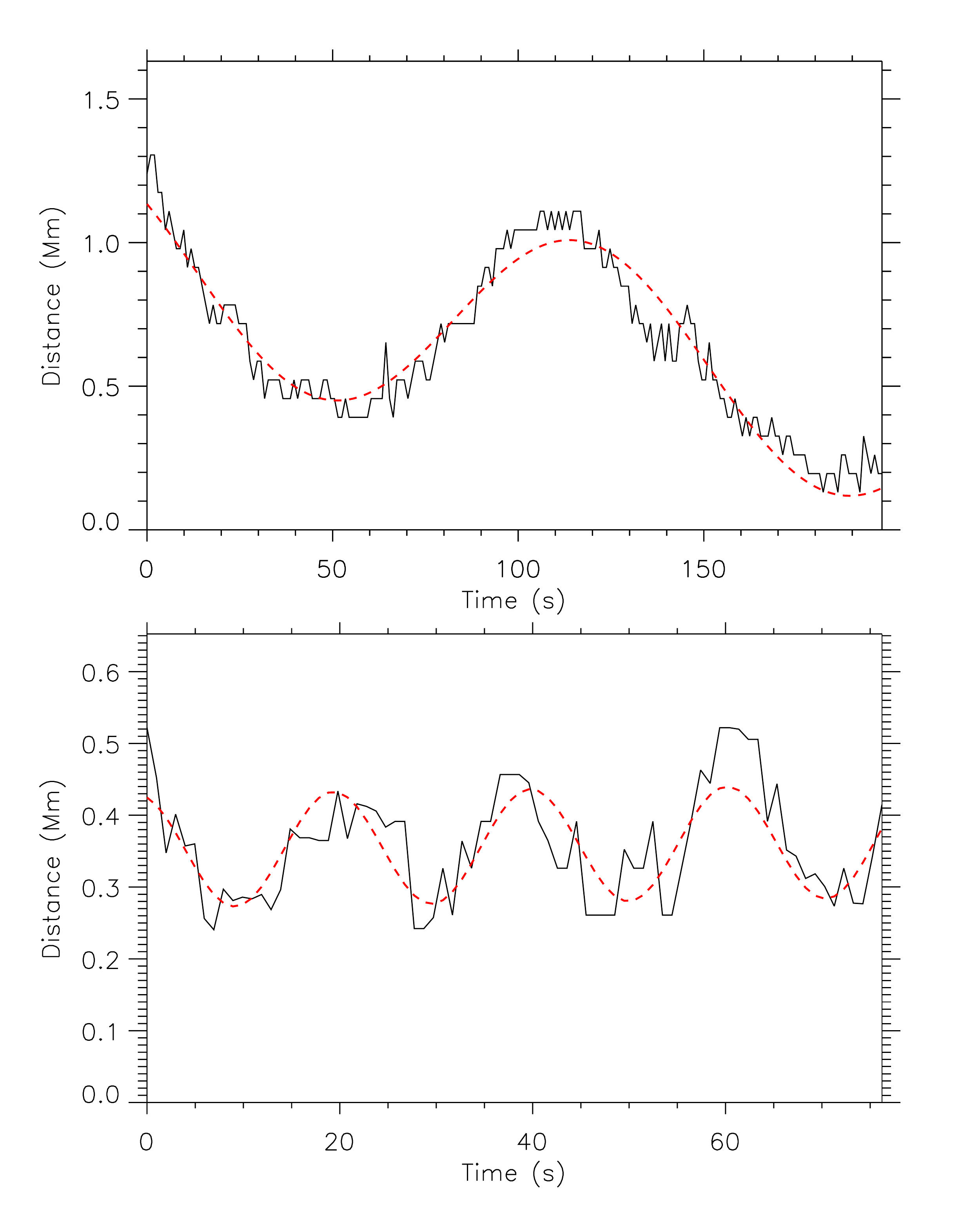}
\caption{The displacement curves, corresponding to two of the waves identified in our dataset, are shown (in their raw form) using the solid black lines. The dashed red lines highlight the fitted waves with properties derived using Fourier analysis. The top panel shows a wave with a period of $\approx138$~s and a displacement amplitude of $\approx358$~km, while the bottom panel shows a wave with a period of $\approx20$~s and a displacement amplitude of $\approx79$~km.} 
\label{fig:fast_slow}
\end{figure}

At each of the five heights considered, over $3000$ spicule features are detected in the time-distance diagrams. Employing Fourier analysis, the properties of the waves present in the transverse motions of these features were determined. As a representative example, the averages and deviations of wave properties found at a height of $6850$~km above the limb are displayed in Figure~\ref{fig:histo}, where the distributions of the displacement amplitudes, periods, and calculated velocity amplitudes of these waves are plotted as histograms. These properties follow  approximate log-normal distributions, which are shown by the solid green lines in Figure~{\ref{fig:histo}}. Log-normal distributions for these properties are consistent with those found in previous studies \citep[e.g.,][]{DePontieu2007, Okamoto2011, Pereira2012}.

\begin{figure}[t]
\centering
\includegraphics[trim=0mm 0mm 0mm 0mm, clip, width=\columnwidth, angle=0]{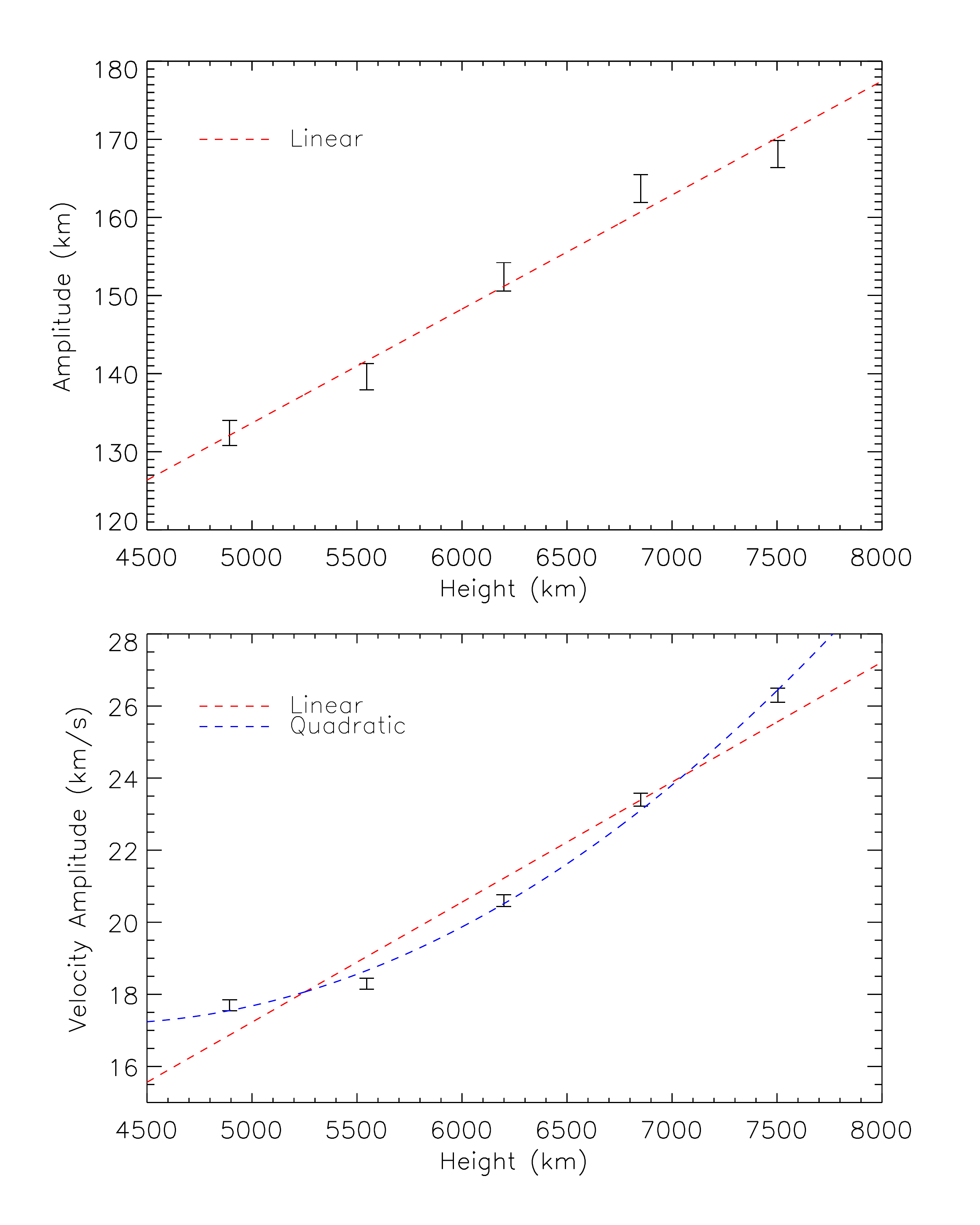}
\caption{Mean values of spicule displacement amplitudes (top panel) and velocity amplitudes (lower panel) plotted against height above the solar limb. Linear lines of best fit are shown in both panels using a dashed red line. In the bottom panel a quadratic fit is shown using a dashed blue line. Errors associated with each data point represent the standard error on the mean.}
\label{fig:amp_h}
\end{figure}

\begin{deluxetable*}{cccccc}[t!]
\tablenum{1}
\tablewidth{0pt}
\tablehead{
\colhead{Height} & \colhead{Number of Waves} & \colhead{Displacement Amplitude} & \colhead{Period} & \colhead{Velocity Amplitude} \vspace{-2mm} \\
\colhead{(km)} & \colhead{} & \colhead{(km)} & \colhead{(s)} & \colhead{(km{\,}s$^{-1}$)}
}
\startdata
4890 & 4880 & 132.4$\pm111.2$ & 55.1$\pm45.0$ & 17.7$\pm10.6$ \\
5550 & 4920 & 139.6$\pm118.8$ & 57.0$\pm48.1$ & 18.3$\pm10.7$ \\
6200 & 5022 & 152.4$\pm128.5$ & 55.8$\pm48.1$ & 20.6$\pm11.6$ \\
6850 & 5209 & 163.7$\pm128.6$ & 53.4$\pm45.4$ & 23.4$\pm13.1$ \\
7500 & 5298 & 168.1$\pm125.4$ & 48.9$\pm39.9$ & 26.3$\pm14.1$ \\
\enddata
\caption{Mean wave properties and their standard deviations at each sampled height. \label{tab:wav}}
\end{deluxetable*}

Average wave properties for each height are presented in Table~\ref{tab:wav}. Importantly, the averages of the displacement and velocity amplitudes appear to be consistent with those found in previous studies of transverse waves in spicules. However, the average period of the waves in the current study are shifted to lower values than those found previously \citep[see the summary provided by][]{Jess2015}. Specifically, the majority of earlier studies found average periods on the order of $80-300$~s \citep[e.g.,][]{1971SoPh...18..403N, DePontieu2007, Kuridze2012, Morton2012, Morton2013, Morton2014}, while we find the average period to be $53\pm45$~s (middle panel of Figure~{\ref{fig:histo}}). It should be noted that the mean has been chosen for comparison with previous studies here. However, as the wave properties approximately follow log-normal distributions, the modal value represents a more useful statistic in understanding the peak of this distribution. As a point of comparison, \citet{1971SoPh...18..403N} found modal and mean periods of $60$~s and $85$~s, respectively, for their observed spicule oscillations, whereas example modal and mean periods found at a height of $6850$~km above the solar limb in this study are $\approx$24~s and $\approx$54~s, respectively, as shown in Figure~\ref{fig:histo}. We consider the detection of these shorter period waves likely due, at least in part, to the unprecedented $\sim1$~s time cadence of the dataset utilized. For comparison, previous investigations using data from the Swedish Solar Telescope provided cadences on the order of $5$~s, which would make it very difficult for the lowest oscillation periods ($<10$~s) identified here to be detected. 

Across all five defined slits, over $16{\,}600$ spicular threads were identified, of which $15{\,}959$ (95.9\%) exhibit at least one complete wave cycle. Of these examples, $8568$ (51.5\%) threads exhibit a single wave, $5770$ (34.7\%) consist of two superposed waves, and $1621$ (9.2\%) have three (or more) superposed waves. These proportions are similar to those found in transverse oscillations of coronal plumes using Auto-NUWT by \citet{Weberg2018}.

Two examples of the identified waves are shown in Figure~\ref{fig:fast_slow}. These are chosen as they have radically different periods and displacement amplitudes, consisting of $\approx$138~s and $\approx$358~km, respectively, for the top panel, with $\approx$20~s and $\approx$79~km, respectively, for the lower panel. Both waves are observed for longer than one full period. The wave identified in the top panel of Figure~{\ref{fig:fast_slow}} has properties consistent with those found in previous studies of transverse spicule oscillations \citep[see the review by][]{Jess2015}, highlighting that these longer period ($>50$~s) waves are also present within our data and are fitted well using our techniques. However, due to the high spatial and temporal resolutions provided by HARDcam, much shorter period waves are able to be identified, including the example shown in the lower panel of Figure~{\ref{fig:fast_slow}}. 

As the wave properties have been determined for each of the five equally-spaced slits above the solar limb, we are able to compare and study characteristics as a function of atmospheric height. The mean values for displacement amplitude and velocity amplitude are shown in Figure~\ref{fig:amp_h}, where both parameters can be seen to increase with height. By fitting a linear line of best fit through the corresponding data points (see the dashed red lines in Figure~{\ref{fig:amp_h}}), the displacement amplitude increases at a rate of $14.6\pm0.8$~km$/$Mm, and the velocity amplitude at $3.33\pm0.08$~km{\,}s$^{-1}/$Mm. The conservation of energy flux requires a reciprocal relation between density and velocity amplitude \citep[see, e.g., the discussions in][]{1974ARA&A..12..407S, 2012Ap&SS.337...33E, 2013A&A...556A.115D, 2015LRSP...12....6K, 2018NatPh..14..480G, 2018ApJ...860...28H, 2020ApJ...892...49H, 2021A&A...648A..77R}. Employing spectropolarimetric inversions of the Ca~{\sc{ii}} spectral line, \citet{2021ApJ...908..168K} revealed evidence that the mass density of spicules decreases exponentially with height, requiring the velocity amplitude to similarly increase to conserve energy flux. Hence, a quadratic fit is presented in the lower panel of Figure~{\ref{fig:amp_h}} using a blue dashed line to show the potential synergy between expected mass density and velocity amplitude. However, we note that this is presented only for completeness, since it is difficult to infer the true nature of the relationship with only five data points. The average periods do not show a similar trend with atmospheric height, instead ranging within the same interval of $48.8-57.0$~s for the five heights considered. 

In order to measure the phase velocity of these waves, it was necessary to identify the same feature across different heights. This was achieved by extracting individual wave properties from a certain atmospheric height and searching through the wave catalog for waves with similar properties identified at an adjacent height. The properties considered for this study were the equilibrium $x$-position of the spicule ($\pm$5~pixels or $\pm$330~km), the midpoint time (whether-or-not the next spicule feature lay between the start and end times of the wave being compared), duration of the oscillation ($\pm50\%$), and the frequency ($\pm10\%$). Based on these criteria, around $140$ waves were found to be suitably similar between each set of adjacent heights, providing large number statistics with a similar proportion (tracked waves in relation to total identified waves) to that documented by \citet{Jafarzadeh2017c}. 

\begin{figure}[t]
\centering
  \includegraphics[trim=0mm 0mm 0mm 0mm, clip, width=\columnwidth, angle=0]{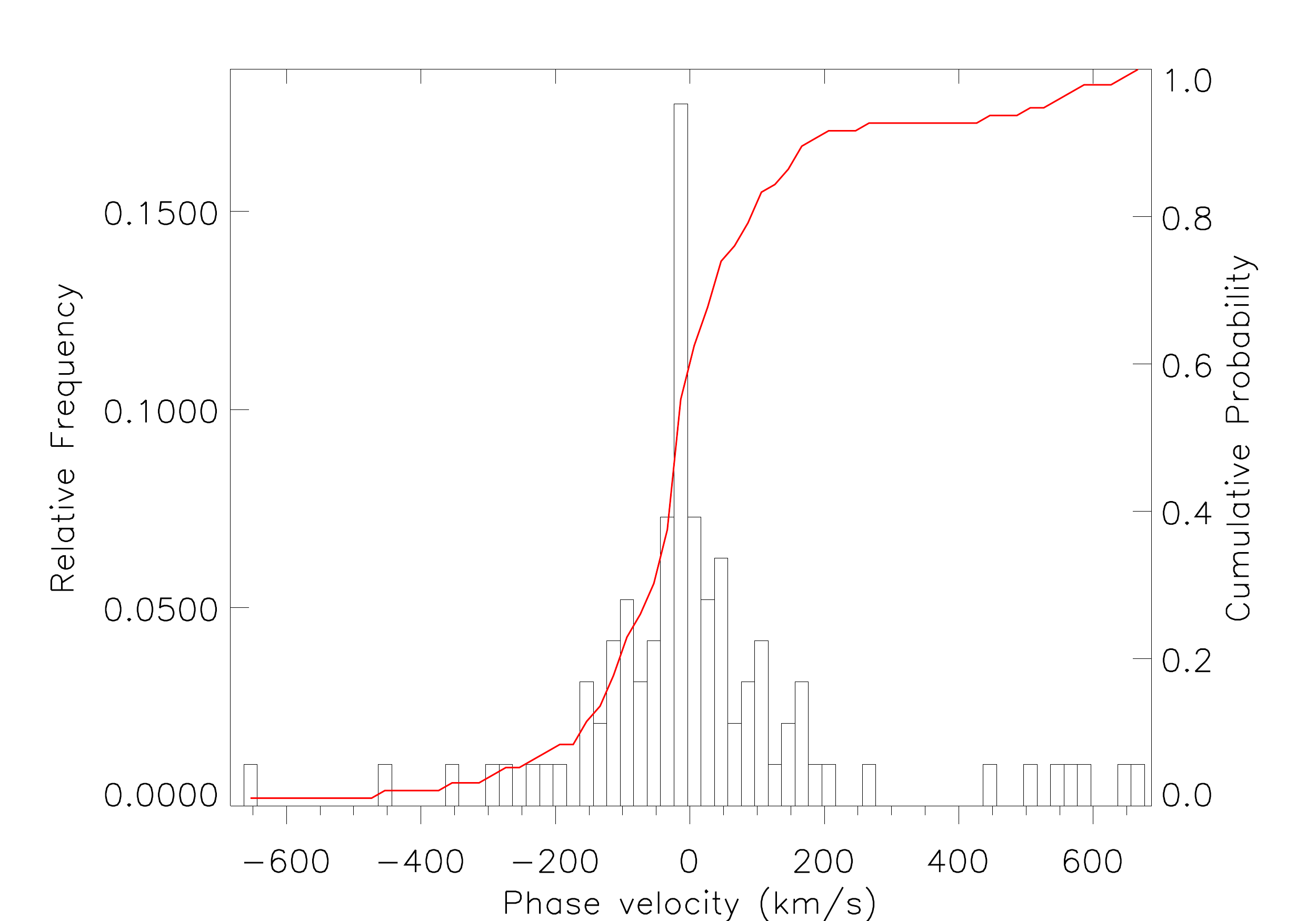}
\caption{A histogram showing the calculated phase velocities of the $135$ propagating waves identified between the heights of $6200$ and $6850$~km above the solar limb. The solid red line denotes the cumulative probability function. A bin width of $20$~km{\,}s$^{-1}$ was used for the creation of this histogram.} 
\label{fig:histo_phv}
\end{figure}

The phase difference between all sets of waves identified at adjacent heights was calculated using Fourier phase lag analysis. The cross-power spectrum was calculated using the representative Fourier spectra of the two waves found at adjacent heights \citep{Bendat2000}. The real part of the cross-power spectrum (co-spectrum) was used to verify that each original Fourier spectrum had a peak at the same frequency. The phase of the cross-power spectrum was then computed at the same frequency to determine the phase lag between the two heights \citep{Vaughan1997}. Finding this phase lag, $\phi$ (in degrees), allows for the calculation of the phase velocity, $v_{ph}$ (in km$/$s),
\begin{equation}
\label{eqn:phase_speed}
    v_{ph} = \frac{360d}{\mathcal{T}\phi} \ ,
\end{equation}
where $d$ is the height difference between the two slits in km, and $\mathcal{T}$ is the period of the wave in seconds \citep{2012ApJ...746..183J, Gonzalez2020}.

Importantly, calculation of the phase velocities of the waves embedded within the spicules allows for the eventual calculation of their energy fluxes. The distribution of phase velocities for the $135$ propagating waves identified traveling between the heights of $6200 \rightarrow 6850$~km are shown in Figure~\ref{fig:histo_phv}, where waves propagating in both the upward and downward directions are identified. Any waves displaying zero phase lags (i.e., providing infinite phase speeds in Equation~{\ref{eqn:phase_speed}}) were classified as standing modes. Due to the relatively small number of standing modes present in our dataset, this type of wave is not included in the histogram depicted in Figure~{\ref{fig:histo_phv}}. We must highlight that the distribution of upwardly and downwardly propagating waves shown in Figure~{\ref{fig:histo_phv}} appear to originate from the same population, with an approximately Gaussian distribution encompassing waves propagating both upwards and downwards. However, when examining the energy flux carried by these waves it is important to examine the direction of energy propagation, which is determined by the sign of their associated phase speed. Thus, distinctions are made between upwardly and downwardly propagating waves for the sake of further analysis, but it must be emphasized that there do not seem to be two distinct populations present in Figure~{\ref{fig:histo_phv}}.

It should be noted that any plasma flows within the spicules will affect the apparent phase velocities of the measured kink oscillations. In the case of upflowing plasma, the apparent (i.e., measured) phase velocities of the upwardly propagating waves will be related to $(v_{ph} + U)$, while the downwardly propagating waves will have apparent phase velocities equal to $(-v_{ph} + U)$, where $v_{ph}$ is the true phase velocity and $U$ is the velocity of the upflow \citep{1995SoPh..159..213N}. This is similar to observations put forward by \citet{2015ApJ...806..132G}, who examined the bulk plasma upflow within a magnetic pore and the subsequent effect this had on the apparent wave speeds of sausage mode oscillations. Strong upflows are typically associated with type~{\sc{ii}} spicules. However, the spicules observed in this study are likely not best characterized by this classification \citep{DePontieu2007a}. This effectively means that the apparent phase velocities of the upwardly propagating waves can be considered an upper limit to their true phase velocities. Conversely, in the case of the downwardly propagating waves, this can be considered as a lower limit. As the velocity of any possible upflows are not known, the measured phase velocities have been used in all further calculations, but it is important to note that this will result in the calculated energy fluxes being upper/lower limits for the upwardly/downwardly propagating kink waves.

Across the four sets of adjacent heights, the occurrence rates of upwardly propagating, downwardly propagating, and standing mode waves were found to be $45\%$, $49\%$, and $6\%$, respectively. This is in contrast to the occurrence rates found by \citet{Okamoto2011} of $59\%$, $21\%$, and $20\%$, respectively. However, the spicules examined by \citet{Okamoto2011} were observed within a coronal hole, so may have different properties to those examined here. Importantly, our present study highlights a more equal balance of upward/downward propagation, with fewer examples categorized as standing modes. The lack of standing mode detections may also be a consequence of the incredibly high spatial and temporal resolutions of the HARDcam dataset, since phase precision is drastically improved as a result of the sub-1~second cadence. 

It might initially be assumed that a roughly equal balance of upward/downward propagation should be expected, due to the high reflection coefficient of the transition region \citep{1982SoPh...75...35H}. \citet{Liu2014} also observed downwardly propagating transverse waves within solar spicules and note that low-frequency (periods of $\approx100$~s) are expected to reflect strongly in the transition region \citep{Suzuki2005}. However, \citet{Okamoto2011} argue that such reflection would result in more standing modes being observed due to the superposition of upwardly and downwardly propagating waves. This would create an imbalance in observations, with more upward than downward propagations detected. This superposition is, however, heavily dependent on the height of the reflecting boundary, the phase velocity of the upward wave, the lifetime of the spicule, and the time that the wave persists for. If there is insufficient time for the reflected wave to interact with the upward wave, due to any combination of the aforementioned criteria, then wave superposition (and hence a standing wave) will not be observed. Although a full characterization of the driving mechanisms behind the downwardly propagating waves, as well as the clear domination of these waves with a phase speed around zero (see Figure~{\ref{fig:histo_phv}}), is beyond the scope of the present work, these are important questions to be investigated in future studies.

\begin{deluxetable}{ccc}[]
    \tablenum{2}

    \tablewidth{0pt}
    \tablehead{
    \colhead{Height} & \colhead{Upward} & \colhead{Downward} \vspace{-2mm} \\
    \colhead{} & \colhead{Phase Velocity} & \colhead{Phase Velocity} \vspace{-2mm} \\
    \colhead{(km)} & \colhead{(km{\,}s$^{-1}$)} & \colhead{(km{\,}s$^{-1}$)}
    }
    \startdata
    $4890\rightarrow5550$ & $128\pm23$ & $75\pm12$ \\
    $5550\rightarrow6200$ & $131\pm23$ & $82\pm23$ \\
    $6200\rightarrow6850$ & $139\pm25$ & $101\pm15$ \\
    $6850\rightarrow7500$ & $147\pm23$ & $128\pm23$ \\
    \enddata
    \caption{Mean phase velocities for each set of adjacent heights that are defined in Table~{\ref{tab:wav}}. \label{tab:ph_v}}
    
\end{deluxetable}

\begin{figure}[t]
\centering
  \includegraphics[trim=0mm 0mm 0mm 0mm, clip, width=\columnwidth, angle=0]{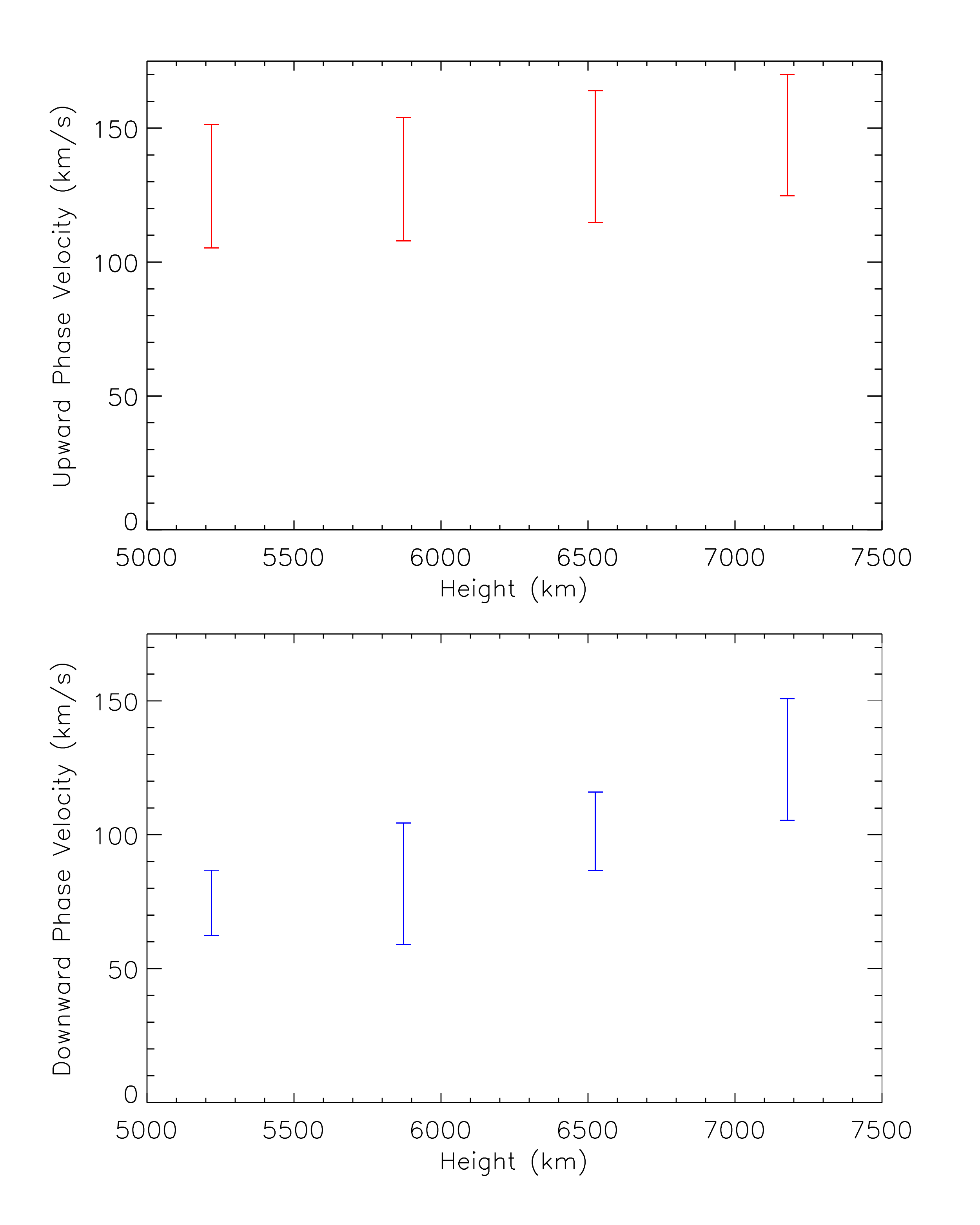}
\caption{Mean values of phase velocity shown with height above the solar limb. The values for upwardly and downwardly propagating waves are shown in the top and bottom graphs respectively. Errors are calculated using bootstrapping.} 
\label{fig:phv_h}
\end{figure}

For each of the four sets of adjacent heights, waves identified as upwardly propagating were segregated from their downwardly propagating counterparts. It was hence possible to calculate the rate of change of phase velocity as a function of atmospheric height for both the upwardly and downwardly propagating waves. Average phase velocities for each set of adjacent heights are shown in Table~\ref{tab:ph_v} and plotted in Figure~\ref{fig:phv_h}, where the upper panel corresponds to the average phase velocity of the upwardly propagating waves, while the lower panel depicts the average phase velocity of the downward propagation. The uncertainties shown in Figure~{\ref{fig:phv_h}} have been calculated following the `bootstrap' methodologies described by \citet{Efron:1979}. Due to the combined presence of traditional (periods $\ge50$~s) and high-frequency (periods $<50$~s) spicule oscillations, it is challenging to assign basic standard errors to the derived phase velocities, especially since the equivalence (or lack thereof) of the driving mechanisms responsible for these characteristics have yet to be observationally and/or theoretically verified. As such, we apply bootstrapping techniques to better constrain the confidence intervals of data following non-normal or unknown distributions \citep[similar to that presented by][]{Simpson:1986,Desmars:2009,Yao:2017}.

The upward phase velocities appear to increase with atmospheric height at a rate of approximately ${10\pm15}$~km{\,}s$^{-1}/$Mm. However, due to the size of the associated uncertainties (see the error bars in the upper panel of Figure~{\ref{fig:phv_h}}), it is difficult to unequivocally stipulate the precise relationship. A more pronounced trend is present in the downward phase velocities (lower panel of Figure~{\ref{fig:phv_h}}), which appear to decrease (as the height sampled decreases) at a rate of approximately $24\pm11$~km{\,}s$^{-1}/$Mm, implying that the wave slows as it travels down the spicule and encounters more dense layers of the lower solar atmosphere.

With the velocity amplitudes and phase velocities of the oscillations measured, it was possible to estimate the energy flux associated with both the upwardly and downwardly propagating waves. In order to calculate the energy flux, a model for the density of the spicules with height is required. \citet{Kuridze2021} observed a limb spicule and derived a model of its density using the Non-LTE Inversion Code using the Lorien Engine \citep[NICOLE;][]{Socas-Navarro2015} inversion code. The final density model takes the form,
\begin{equation}
    \rho(y) = \rho_{0}e^{(y-h_{0})/\Lambda} \ ,
\end{equation}
where $y$ is the height above the solar limb, $\rho(y)$ is the spicule density as a function of height, $h_{0}$ is the base height, $\rho_{0}$ is the density at the base height, and $\Lambda$ is the density scale height. Values for our energy flux calculations were taken directly from \citet{Kuridze2021}, where $\rho_{0}\approx6\times10^{-7}$~kg{\,}m$^{-3}$, $h_{0}=2000$~km, and $\Lambda = 1500$~km.

The energy flux, $F$, from transverse waves in a multiple flux tube system can be calculated as, 
\begin{equation}
    F \approx f\frac{1}{2}(\rho_{i}+\rho_{e})v^{2}v_{gr} \ ,
\end{equation}
where $f$ is the density filling factor, $\rho_{i}$ is the density inside the flux tube filled in by the spicule, $\rho_{e}$ is the density outside the spicule, $v$ is the velocity amplitude, and $v_{gr}$ is the group speed \citep{VanDoorsselaere2014}. For propagating kink waves, the group velocity can be approximated by the phase speed, $v_{ph}$, as they are only weakly dispersive \citep{Terradas2010,2021SSRv..217...73N}. The internal density for spicules can be assumed to be much larger than the external density, i.e., $\rho_{i}\gg\rho_{e}$ \citep{Uchida1961}, providing a simplified equation for the energy flux,
\begin{equation}
    F \approx f\frac{1}{2}\rho_{i}v^{2}v_{ph} \ .
    \label{eqn:flux}
\end{equation}

\begin{figure}[t]
\centering
  \includegraphics[trim=0mm 0mm 0mm 0mm, clip, width=\columnwidth, angle=0]{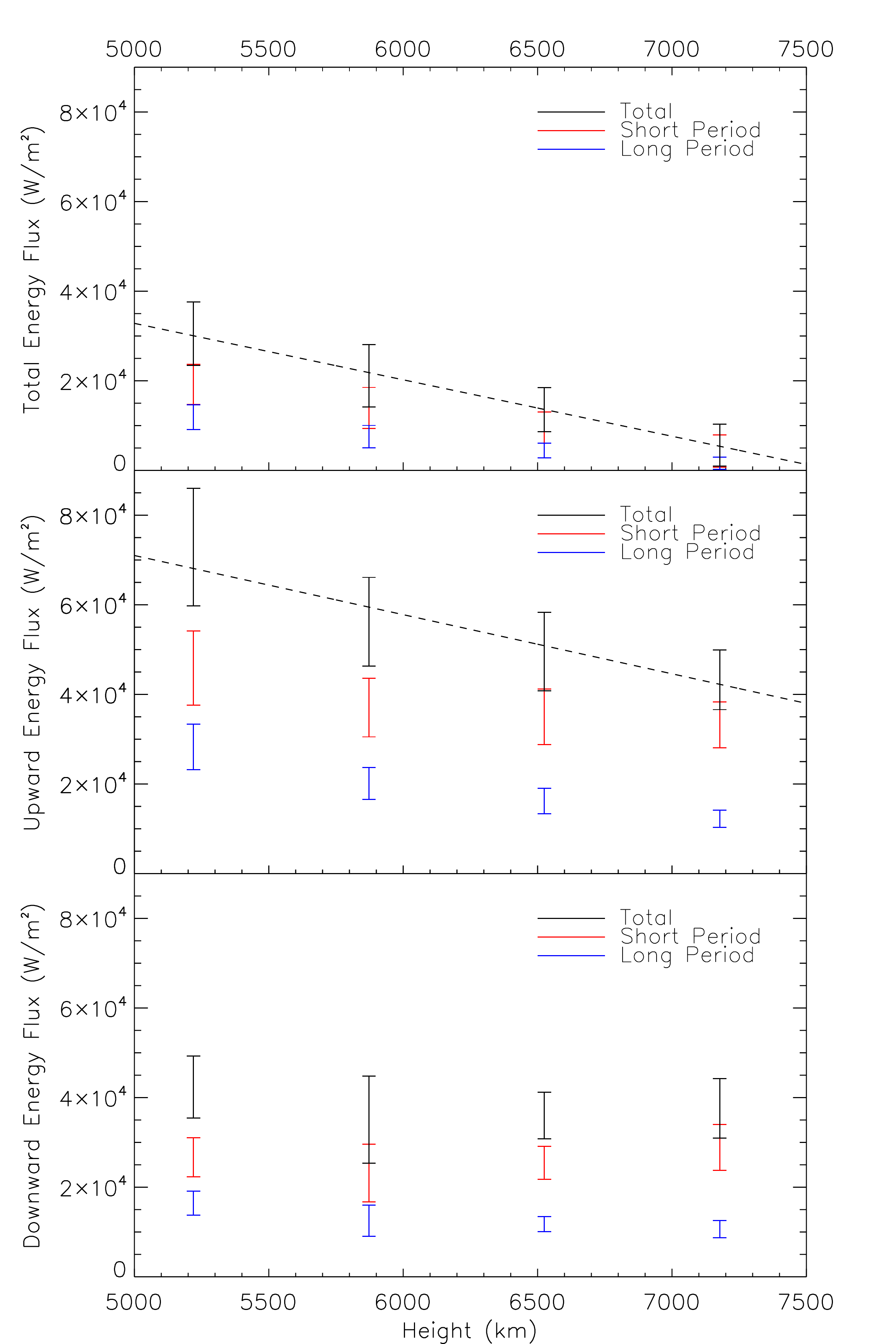}
\caption{Energy flux estimations as a function of atmospheric height for all propagating waves (upper panel), upwardly propagating waves (middle panel), and downwardly propagating waves (lower panel). The total energy flux provided by short/long period waves is shown in black, while the energy fluxes for short- ($<50$~s) and long-period ($>50$~s) waves are shown in red and blue, respectively. The energy fluxes provided by the full set of waves (including upwardly and downwardly propagating) and for all upwardly propagating waves are depicted, using a linear line of best fit, as a dashed black line in the upper and middle panels, with gradients equal to $-12{\,}600$~W{\,}m$^{-2}/$Mm and $-13{\,}200$~W{\,}m$^{-2}/$Mm, respectively. } 
\label{fig:flux_h}
\end{figure}

Taking the upper limit of the spicule density filling factor as $5\%$ \citep{Morton2012} allowed the energy fluxes to be calculated for each adjacent set of heights, which are displayed individually for all propagating waves (top panel) alongside upwardly (middle) and downwardly (bottom) propagating waves in Figure~\ref{fig:flux_h}. For all waves examined, it can clearly be seen that there is a decrease in energy flux with height, indicated using solid black data points in the upper panel of Figure~{\ref{fig:flux_h}}. A linear line of best fit is presented using a dashed black line in the upper panel of Figure~{\ref{fig:flux_h}}, with a gradient of $-12{\,}600$~W{\,}m$^{-2}/$Mm. However, an exponential fit would perhaps be more appropriate, since the main factor for the energy flux decrease is expected to be density stratification, which is typically represented by a decaying exponential profile with height \citep[e.g.,][]{Verth2011}. Due to the relatively small number of data points under consideration, a linear fit has been chosen for simplicity. Regardless of the fitting function employed, the important message is that the energy flux of the propagating transverse waves clearly decreases with atmospheric height, hinting at some sort of damping and/or dissipation process. 

It is important to consider the effect of using a filling factor of $5\%$. This means that Equation~\ref{eqn:flux} estimates the energy flux under the assumption that the waves are omnipresent, i.e., does not take into account the sporadic nature of the observed wave motion. In addition, as the waves are not seen to exist in all spicules, the actual filling factor, $f$, should be reduced to account for this effect. Thus, the estimation based on Equation~\ref{eqn:flux} gives us the upper limit of the energy flux in the waves. However, as the filling factor is a multiplicative term, this only affects the magnitude of any energy flux estimations. The trends in energy flux examined with respect to height are independent of any adjustment to the filling factor. For example, using the relatively low filling factor of $0.6\%$ suggested by \citet{1972ARA&A..10...73B} will simply lower all energy flux and rate of change of flux values by a linear factor of $0.12$ when compared to those values calculated with a filling factor of $5\%$. The values presented in the text and within Figure~{\ref{fig:flux_h}} utilize a filling factor of $5\%$, unless stated otherwise, and should therefore be taken as upper limits.

For all upwardly propagating waves, we observe the energy flux to decrease as a function of height at a rate of $-13{\,}200\pm6500$~W{\,}m$^{-2}/$Mm, which is indicated in the middle panel of Figure~{\ref{fig:flux_h}} using a dashed black line derived from a linear least-squares fit.  For completeness, it is estimated that energy fluxes in the range of $10^{3}-10^{4}$~W{\,}m$^{-2}$ are required to heat the chromosphere \citep{Withbroe1977}. Hence, the total energy flux, in addition to the measured rate of energy flux decay with height, are on the same order as the total energy input required to provide basal heating to the solar chromosphere. Even considering the relatively low filling factor of $0.6\%$, as suggested by \citet{1972ARA&A..10...73B}, the rate of energy flux decrease would be $-1580\pm780$~W{\,}m$^{-2}/$Mm, still within the range that is needed to balance the radiative losses of the chromosphere. By contrast, the energy flux for all of the waves propagating downwards does not appear to depend on the height sampled (black data points in the lower panel of Figure~{\ref{fig:flux_h}}), with the energy flux estimates remaining consistent (${\sim4\times10^4}$~W{\,}m$^{-2}$) across the height range of approximately $7500 \rightarrow 4900$~km above the solar limb. The similarity in the rate of energy flux drop off in height is consistent between the full set of waves (upper panel of Figure~{\ref{fig:flux_h}}) and the upwardly propagating ones (middle panel of Figure~{\ref{fig:flux_h}}). This is to be expected, since the downward energy flux remains approximately constant with atmospheric height.

The decrease in upward energy flux with atmospheric height may be due to at least three different factors: (1) physical thermalization of wave energy into localized heat via dissipation mechanisms \citep[e.g.,][to name but a few examples]{Hollweg1986, 2009A&A...497..525H, Antolin2015, Antolin2018, Okamoto2015}, (2) damping of detectable transverse waves through the process of mode conversion, where kink mode amplitudes decay as a result of the transfer of energy from transverse kink oscillations to azimuthal Alfv{\'{e}}n motions \citep{Pascoe2010, Pascoe2012, Pascoe2013}, and/or (3) reflection of the waves downward at varying heights above the solar limb \citep{1982SoPh...75...35H, Suzuki2005}. Tentative observational evidence has shown that torsional Alfv{\'{e}}n and kink waves may exist concurrently in spicules, providing credence for the applicability of mode conversion processes \citep{DePontieu2012}. Previous modeling work by \citet{1984ApJ...285..843S} has shown that Alfv{\'{e}}n waves within spicules can produce high-frequency signatures, including periodicities of $112$, $37$, and $22$~s for the fundamental, first, and second harmonic resonant periods, respectively, which are similar to the periodicities found in our current work. Employing simultaneous plane-of-sky imaging and line-of-sight Doppler measurements will allow more precise definitions of the embedded spicule wave modes, which will allow the high-frequency Alfv{\'{e}}n modes to be examined and compared to the models put forward by \citet{1984ApJ...285..843S}.

In order to establish if the wave energy is dissipated in the form of localized heating, measurements of thermal processes in the vicinity of these spicules are necessary. This may be achieved using differential emission measures of optically thin coronal EUV observations directly above the spicules \citep{McIntosh2012, 2012SoPh..280..425V}. An alternative approach would be to use the Atacama Large Millimeter/Submillimeter Array \citep[ALMA;][]{Wootten2009, 2016SSRv..200....1W} to find the temperature of the spicules and the surrounding plasma \citep{Chintzoglou2021, Henriques2021, Jafarzadeh2021}. Importantly, the timing information related to the decay of the spicule oscillations would need to be harnessed to provide both spatial and temporal information to examine localized temperature fluctuations that may be a result of thermalization mechanisms. While this is beyond the scope of the present work, it will form the basis of a follow-up study over the coming months.

The downwardly propagating waves maintain an approximately constant energy flux through a reduction in both velocity amplitude and phase velocity as they travel down the spicule, visible in Figures~\ref{fig:amp_h} and \ref{fig:phv_h}, respectively. It is likely that this is due to the wave interacting with the denser plasma at lower heights above the solar limb, resulting in a slower Alfv{\'{e}}n speed in these regions \citep{Okamoto2011}. This is not unexpected, as the theoretical modeling of propagating kink waves in longitudinally stratified waveguides found that phase velocities and velocity amplitudes decrease with height \citep{2011ApJ...736...10S}. 

It has been proposed that in order to supply the quasi-steady effects needed to heat the solar atmosphere, the dissipation of short period waves is of paramount importance \citep{2005ApJ...631.1270H,2008ApJ...680.1542H,2011ApJ...736....3V}. The energy flux carried by both short period ($<50$~s) and long period ($\ge50$~s) waves between each set of adjacent heights is shown in Figure~\ref{fig:flux_h} using red and blue data points, respectively. In order to calculate the associated energy flux for the propagating wave modes, new filling factors were calculated by combining the previously used spicule density filling factor \citep[$5\%$;][]{Morton2012} with the fraction of waves which were found to fall into each relevant category (i.e., $<50$~s or $\ge50$~s). The new filling factors were approximately $2.5\%$, which is a result of the $50$~s boundary being very close to the average period found at each height (see Table~\ref{tab:wav}). 

It can be seen from Figure~\ref{fig:flux_h} that the energy flux of the short period waves is greater than that of the long period waves for the full set of propagating waves (upper panel), and both the upwardly propagating (middle panel) and downwardly propagating (lower panel) waves. For the full set of propagating waves and the upwardly propagating waves, both the short and long period waves show a similar energy flux decrease with height as that for the total energy flux values. The energy flux of both the short and long period downwardly propagating waves show a similar lack of dependence on atmospheric height, which is consistent with the total energy flux measurements. This suggests that both short and long period upwardly propagating waves have the potential to heat the solar atmosphere, although the short period waves have a larger energy flux across all heights, giving them a greater potential capacity for heating.

\section{Conclusions} 
\label{sec:conc}
The results presented here represent a sizable increase in the statistical population of examined transverse spicule oscillations. Our use of data with a time cadence of $\sim1$~s also allowed for the identification of high frequency waves, similar to those found by \citet{Okamoto2011}, with periods as short as $10-20$~s, only now with a significant increase in the examined population size. Observations with even higher spatial and temporal resolutions may allow for the detection of even shorter period and smaller-scale oscillations, and further extend the statistical distributions (see, e.g., Figure~{\ref{fig:histo}}) down to even smaller values. 

We examined the wave properties of spicule oscillations across multiple atmospheric heights, which facilitated the calculation of associated phase speeds, hence allowing us to categorize the waves as either being upwardly/downwardly propagating or standing. Almost an equal balance was found between upwardly (45\%) and downwardly (49\%) propagating waves, in contrast to the earlier study by \citet{Okamoto2011}, who found that upwardly propagating waves were dominant in their time series. However, the observations presented here are in close proximity to the solar active region NOAA~AR12391 and may therefore have distinctly different properties to the coronal hole observations examined by \citet{Okamoto2011}. 

Directional information for the spicule waves allowed the calculation of their associated energy flux as a function of upwardly and downwardly propagating waves across a number of atmospheric heights. Energy flux estimates are relatively consistent across all heights for the waves propagating in a downwards direction. However, for the upwardly propagating waves, a negative correlation with height is demonstrated, with the overall energy flux decreasing at a rate of $-13{\,}200\pm6500$~W{\,}m$^{-2}$/Mm calculated with a spicule filling factor of $5\%$ (or at a rate of $-1580\pm780$~W{\,}m$^{-2}/$Mm using a lower-limit filling factor of $0.6\%$). The mechanism responsible may either be due to thermalization of the mechanical wave energy or mode coupling, although investigation of the proportional contributions of each mechanism are beyond the scope of this study. If even a small fraction of the wave energy carried in the transverse waves of the spicules examined is deposited as thermal energy, then it may significantly contribute to the $10^3 - 10^4$~W{\,}m$^{-2}$ requirements needed to balance the radiative losses of the chromosphere \citep{Withbroe1977}.


\begin{acknowledgments}
WB, DBJ, and FPK acknowledge support from the Leverhulme Trust via the Research Project Grant RPG-2019-371.
DBJ and SDTG wish to thank Invest NI and Randox Laboratories Ltd. for the award of a Research \& Development Grant (059RDEN-1), in addition to the UK Science and Technology Facilities Council (STFC) for the consolidated grant ST/T00021X/1. 
DBJ also acknowledges funding from the UK Space Agency via the National Space Technology Programme (grant SSc-009).
VMN acknowledges support from STFC grant ST/T000252/1 and Russian Foundation for Basic Research (RFBR) grant (grant No. 18-29-21016).
SJ acknowledges support from the European Research Council under the European Union Horizon 2020 research and innovation program (grant agreement No. 682462) and from the Research Council of Norway through its Centres of Excellence scheme (project No. 262622). 
The Dunn Solar Telescope at Sacramento Peak/NM was operated by the National Solar Observatory (NSO). NSO is operated by the Association of Universities for Research in Astronomy (AURA), Inc., under cooperative agreement with the National Science Foundation (NSF). 
The authors wish to acknowledge scientific discussions with the Waves in the Lower Solar Atmosphere (WaLSA; \href{www.WaLSA.team}{www.WaLSA.team}) team, which is supported by the Research Council of Norway (project number 262622), and The Royal Society through the award of funding to host the Theo Murphy Discussion Meeting ``High-resolution wave dynamics in the lower solar atmosphere'' (grant Hooke18b/SCTM). 
\end{acknowledgments}

%

\vspace{5mm}
\facilities{DST \citep[HARDcam;][]{2012ApJ...757..160J}}


\software{MGN \citep{Morgan2014}, Auto-NUWT \citep{Morton2013, Weberg2018}}



\bibliography{ApJbib}{}
\bibliographystyle{aasjournal}

\end{document}